\documentclass[10pt,a4]{article}
\usepackage{latexsym}
\usepackage{amsfonts}
\usepackage{array}
\usepackage[english]{babel}
\usepackage{amsmath}
\usepackage{amssymb}
\usepackage{amsthm}
\usepackage{latexsym}
\usepackage{graphicx}

\def\bE{\begin{equation}}
\def\eE{\end{equation}}
\def\bEA{\begin{eqnarray}}
\def\eEA{\end{eqnarray}}
\def\bEAnn{\begin{eqnarray*}}
\def\eEAnn{\end{eqnarray*}}
\begin{document}
\title{{\huge \sc Eco - No(?) - Physics}\\ - comments and reflexions - }
\author{Stefan Reimann\footnote{Contact address: sreimann@isb.unizh.ch}\\
Swiss Banking Institute\\ University of Zurich}
\maketitle
\begin{abstract}
{\em How can econophysics contribute to economics? Since the relation between basic principles of physics and economics is not established, there is no reason why physical theories should be of any value for economic theory. While economic theories leave the physicists largely without orientation in this field, econo-physicists should orient themselves at concrete problems from economic practice rather than from economic academics. Thereby physicists should respect the lead of economists. This then also puts physics closer to econometrics. Then the natural strength of physics in dealing with empirical data as the fundamental basis for its research could be valuable for both, theory as well as for applications in economics.}  
\end{abstract}

\noindent  
The term Econophysics was coined by E. Stanley to give a name to a recently born academic area whose concern is to apply methods and tools 
from statistical physics, particularly the more general theory of multi particle systems, and the theory of complex systems to economic data, including non-equilibrium concepts due to turbulence and multifractality. Typical objects of econophysics are statistical properties of price fluctuations, correlations, or the scaling behavior of distributions. Main emphasis is on empirical data. ''{A new field - econophysics - opened with the expectations that the proven methods of the physical sciences and the newly born science of complex systems could be applied with benefit to economics. .. ; while the scope of classical econometrics is limited to dynamical models of time series, econophysics uses all tools of statistical physics and complex systems analysis, including the theory of interacting multi agent systems."} (Focardi \& Fabozzi, 2004). For overviews about related concepts and results see the monographs by Mantegna and Stanley (2000), Bouchaud and Potters (2000), Paul and Baschnagel (2000), and Voit (2005). \\

From a theoretical viewpoint, the 'merger of economics and physics', as suggested by the term 'econophysics', would benefit if basic principles of the two sciences could be shown to be intimately linked with each other. Without uncovering or establishing a close relationship between 'fundamental principles' of physics and economics, respectively, the relation is left to the level of verbal and metaphorical analogies. like 'the center of gravity'. This already raises a first serious problem: While physics have revealed sound First Principles, the collection of First Principles of economics and their formulation might not be beyond controversy between different schools of thought in economics. Hence, which principles should be considered? In the following I will mention two examples, one is the possible relation between principle of utility maximization and the Principle of Least Action, while the other one concerns the concept of Entropy which may be related to the Efficient Market Hypothesis. \\

The {Hamiltonian principle of Least Action} is at the very bottom of physics. In economic terms it might be rephrased as: 'Among all possible trading processes those are realized for which the corresponding time aggregated value process has an extremum.' This, in fact, sounds similar to the idea that trading activities are governed by the investors' goals to maximize their utility over some period. A conjecture therefore is that the principles of {\em Least Action} and {\em Utility Maximization} are intimately linked to each other.  Another topic that might serve to establish a closer relation between both disciplines concerns the term {\em entropy} which, roughly speaking, might be regarded as a measure for uncertainty. According to this viewpoint, trading activities are such that uncertainty is maximized since otherwise, if there would be some piece of information on the market, it would be used for profit and hence will be removed. This analogy would link the {\em Principle of Maximum Entropy} or, equivalently of {\em Minimal Information} to the {\em Efficient Market Hypothesis}, see Gulko (1999). While this verbal analogy sounds promising, it has serious conceptual problems: While in a closed thermodynamic system, entropy approaches its maximum, it is not clear that a market is a closed system. Even more, besides the Shannon-Weaver Entropy, in physics called the Boltzmann-Gibbs Entropy, there are other entropy measures due to A. Renyi or E. Tsallis. The question therefore is about which entropy measure is appropriate for a financial market, if any of these. 
On the other hand, the principle of maximum entropy applied to prices appears to give a a promising alternative to option pricing, given that there is some agreement about which entropy is adequate (Buchen \& Kelly (1994)). Nonetheless the general question whether entropy and related concepts play a role in economic systems compatible to their role in physical systems remains open, but, in my opinion, deserves careful further attention. \\

Physics and economy have diffenent cultural roots. Economics has evolved not from a natural science rather than from social sciences through a line of deep intuitions into different and sometime mutually contradictory schools of thought. Therefore, 'econophysics' is not an interdisciplinary field rather than should be regarded as a {\em transcultural hybrid}. This also becomes obvious if one looks at the respective aims of the two disciplines: While physics aims to understand the dynamics of a system, economics seems to be more interested in developing applications, from the development and pricing of financial instruments to questions concerning management and political issues. In this respect, economics is closer to engineering sciences. A better understanding of the economic system under consideration might be helpful for inventing good applications, but for various purposes a good description suffices. If so then there is no need for an explanatory model. This is what econometrics gains its high reputation in economics for.\\

Interestingly the relation of economics and empirical data has been a problematic one for quite a long time. Empirical research on price fluctuations tracing back to Louis Bachelier's work in 1900, such as work by Holbrook Working (1934), Alfred Cowles (1933,1937), and Maurice G. Kendall (1953), was largely ignored by the economic community for about 60 (!) years. 
As the distinguished Cambridge economist N. Kaldor (1972) put it: {" In economics, observations which contradict the basic hypotheses of the prevailing theory are generally ignored. ... And
where empirical material is bought into conjunction with a theoretic model, as in econometrics, the role of empirical
estimation is to illustrate or to decorate the theory, not to provide support to the basic hypothesis."} Physics as an empirically based science therefore did not have any place in economics in these times.\\

This is different with mathematics.
Since about 1950 economic thinking has become dominated by mathematical reasoning originating from Arrow and Debreu's {General Equilibrium Theory}. 
As indicated by the subtitle, the aim of the {\sc Theory of Value} is to {\em explain} prices in the sense of {\em logically deducing} statements from a minimal set of assumptions - regarded as axioms. It should be noted that mathematics is concerned with consistency of formal statements rather than with empirical 'truth' , i.e. the match between a theory or a model and empirical data. Consequently research in these times was exclusively concerned with clarifying minimum requirements in
terms of "basic assumptions" rather than with empirical investigations about the reliability of the assumptions or the relevance of results in relation to actual prices. The following citation by A. Einstein (1954) may throw a spotlight on the different perspectives of economics and physics: {\em Given a set of axioms, ... the skeptic will say: 'it may well be true that a system of equations is
reasonable from a logical standpoint. But it does not prove that it corresponds to nature'. Your are right, dear skeptic. Experience
alone can decide on truth} \\

Data appear to be a natural field in which physics can find its place being valuable for economics by testing the reliability of a model and its assumptions as well as by comparing its predictions with empirical data. The point here is that the multiplicity of theoretical models in economics leaves the physicist without clear orientation. In the following I will consider only three prominent economic examples: the General Equilibrium Theory, the Capital Asset Pricing Model, and the Black and Scholes model for option pricing.\\

  'A theory must have predictive power and has to be falsifiable.'
From this point of view the General Equilibrium Theory appears to be weak. The theory actually boils down to the statement that the excess demand is a continuous vector field on a smooth manifold of prices (this is actually the Walras law) obeying some additional boundedness conditions. Without strong additional but economically implausible constraints there is typically a set of equilibria (Hildenbrand and Kirman, 1988). Since this theory is essentially static, no statement is involved about which of these equilibria is selected during dynamics. Furthermore, the vector field representing the excess demand, can be continuously deformed such that an arbitrary equilibrium price (singular point of the vector field) can be obtained. This has been actually known since the work by Mantel - and Sonnenschein in the 1970-th under the name: 'Anything Goes Theorems'. Taking all this together, this theory lacks explanatory power in that it is not falsifiable. An important basic assumption of the theory is that agents are independent from each other rather than 'interact'. This assumption clearly is unrealistic. No one will deny that agents' actions and preferences are not independent from each other. The physicist should keep in mind that 'interaction' in economics is more general than only exchanging 'particles'. It can also mean that the agents' preferences are not independent anymore from each other but correlated. Thus, if one allows the economic agents to interact, then, as F\"ollmer showed, that there might be no market clearing price and hence no equilibrium price at all. His result (Foellmer, 1974) is essentially borrowed from the field of interacting particle systems. In the light of this result, empirical facts seriously challenges the applicability of this theory from a real world perspective.\\

An other theory provided by economists some time ago is the Capital Asset Pricing Model (CAPM), which is a pillar of Modern Portfolio Theory. This pricing theory relies essentially on the assumption that investors master uncertainty by considering only the first two moments of the respective distribution. Investigations of empirical return distributions even quite some time ago have instead shown that empirical distribution are non Gaussian rather than exhibit fat tails. Thus rare events are not taken into account by the investment decisions based on the CAPM. From the Gaussian assumption it moreover follows that only the first two moments enter the theory. Particularly the $\beta$'s are a simple function of the co-variances which, in this context, are thought to carry essential information about assets and their risk. Recent investigations of {\em empirical} correlation matrices have revealed that they exhibit many properties that are equal to those of a certain class of random matrices, the so-called Gaussian Orthogonal Ensemble. This approach traces back to E. Wigner and his work about the spectra of heavy nuclei. From a statistical point of view therefore empirical correlation matrices carry only a very small portion of statistically significant 'information' about the financial market.
While in the stylized world of theory and classrooms the CAPM works nicely, it is of minor value from an empirical, i.e. practical viewpoint.\\

A third candidate for an economic theory that might provide some orientation for physicists is the option pricing formula provided by the works of Black and Scholes and Merton. It is esentially based on a 'no-arbitrage' argument and also assumes, in its standard version, that the world is Gaussian. This pricing model was celebrated in its beginning for its normative power, not its explanatory power. It  did not explain option prices but gave a rule for pricing options. As shown by Longstaff and Rubinstein in 1994, for literature see Buchen \& Kelly (1999), market option prices since 1987 are inconsistent with the traditional Black - Scholes formula. Empirical corrections of the theory had to be included like adjusting for stochastic volatilities, giving rise to what usually is called the 'volatility smile'. But recall, the volatility smile simply states that the model is incorrect. However, recent work by Borland (2002) and others has generalized the standard Black-Scholes model particularly by considering non-Gaussian distribution. The volatility smile in her theory comes from the theory itself and thus is not an heuristic correction. Interestingly, the distributions used are in fact exactly those which maximize the Renyi- and the Tsallis entropy. \\

In my opinion, (at least most of) economic theories therefore leave the physicist without any reliable orientation. What then can a physicist rely on? Without an established relation between the fundamentals of physics and economics he should prevent himself from recycling successful theories and models from his field. For example, while Ising-like or glassy systems look appealing to model interacting particle systems, the problem is that one has no information about, for example, what the nature of the interactions is. Analogies with other field theories have the danger to stay at a purely verbal or formal level without providing the opportunity to confront the theory with hard data.\\

In this situation, I think, a physicist should better stay with data on a purely descriptive level, the aim being to extract regularities from the data. Such properties have been 
called Stylized Facts, while physicist might call them 'invariant properties', since these are statistical properties that are (almost) invariant under the choice of the market considered. About stylized facts and a careful discussion also of methods see Cont (2000). The idea is that these regularities might serve as a basis for first steps towards a theoretical understanding. Unfortunately the collection of stylized facts known today is not sufficient to single out a unique theoretical basis. An other problem consists in the fact that, unlike to physical practice, a physicist can not perform sound experiments on a financial market. An experiment, injecting money for example, will irreversibly change the market. Physicists should be aware of that this fact distinguishes an economic system significantly from (typical) physical systems. As a consequence repeated experiments under identical conditions are impossible. Thus he is left to observations. Economists have of course been aware of this fact for a long time. This might explain economists' strategy to have given theories, mathematical statements as well as 'economic stories' upper priority than data for a long time. \\

Having a theory in mind has the intrinsic danger to destroy an unbiased view on data. There is a prominent example from physics itself: The astronomer Kopernikus observed the trajectories of heavy bodies around the sun. His motivation was to quantitatively show that the trajectories come from the rotation of Platonian Solids - the sign of an god-given harmony of the universe. He gave this 'believe' priority over what he in fact observed and finally started to manipulate data to make them fit into his view. It was then Newton's merit to be able to find his law of gravitation from these 'intentional' data. In general, if one has a theory in mind, there is always the danger of selective data collection. The mismatch between data and expected outcome of a believed theory often is called a 'puzzle' in economics. But then, how to work with data? \\

Proposing a model and then selecting data which fit to the model was already criticized by Kaldor. Practitioners are aware of many problems and questions arising from the real world. These questions and concrete problems may serve as guidances for physicists. The advantage would be that then physicists' efforts were intimately related to practice, opening acceptance of their research by economic practice, and, furthermore, could be a step towards exemplary problem solving. In other words, my proposal is to solve selected problems rather than to set up formal models. Obviously, if the problem concerns an economic question, these problem should come from economists and particularly economic practice rather than from academia. Practitioners thus may be the people from whom physicists can learn what relevant problems are - from an application's and also from a theoretical point of view. In this case, physicists should be honest enough and accept the lead of economists in that respect. But here comes another problem: The definition of a problem includes the definition of the methods used to at least discuss the problem. While it is difficult to 'define' a problem {\em per se}, it is even more difficult to concretely describe a problem in a way such that is it accessible to a foreign discipline and its methods. It is certainly very hard and requires lot of patience and mutual respect to work together on formulating questions and finally defining concrete problems to work on. 
 
\section*{Summary}

In the last decades, physicists have become attacted by problems rising from economics and particularly from finance (Farmer, 2005). Without having established the congruence between basic principles between physics and economics, there is no 'econophysics' as a sound scientific area. Econophysics is more than methods and rephrasing successful physical models in financial terms. The dominance of mathematical modeling and story telling in the field of economics has often lead to a culture of formal or verbal analogies with minor predictive power. This style has prevented most of previous economic research from empirical investigations. Without a sound theoretical fundamental basis in economics and without elaborated empirical investigations of empirical data, physicists are left without any clear orientation in this field.  I am convinced that both fields can gain a lot from each other while they have to respect their fundamental differences. \\

A valuable path for physics in the field of economics might be to solve concrete problems from economic practice and then to move towards the stage of exemplary problem solving. This means to give highest priority to empirical data and to a sound formulation of concrete problems. These problems should come from practice rather than from economic theory. This then may open an opportunity for a collaboration which is potentially valuable for both sides. Recent activities in the field such as the successful Capital Mutual Fund founded by J.-Ph. Bouchaud and M. Potters show that a careful and deep application of methods and tools from physics can earn money, while it additionally provides deep insides into directions of theoretical understanding. This could serve as a hint about a valuable direction to go.  

\section*{References}
{\small
\begin{enumerate}
\item Borland, L., 2002. A theory of non-Gaussian option pricing, Quantitative Finance Vol 2(6), pp 415-431\par
\item Bouchaud, J.-Ph., Potters, M., 2000. Theory of Financial Risk - From Data
Analysis to Risk Management, Cambridge University Press, Cambridge MA\par
\item Buchen, P.W., Kelly, M., 1996. The maximum entropy distributioon of an asset inferered from option prices. The Journal of Financial and Quantitative Analysis, Vol. 31, No. 1, pp 143-159\par
\item Cont, R., 2001. Empirical properties of asset returns: stylized facts and statistical
issues. Quantitative Finance, Vol. 1, 223-236\par
\item Einstein, A., 1954. Ideas and Opinions, Based on {\em Mein Weltbild} (1934, 1955), ed C. Seelig, transl. S. Bargmann, Wings Books, New York \par
\item Farmer, D., 1999. Physicists attempt to scale the ivory towers of finance, Computating in Science and Engineering, Vol 1(6), pp 26-39\par
\item Focardi, S. M., Fabozzi,  F. J., 2004. The Mathematics of Financial Modelling \& Investment Management. Wiley, p. 76\par
\item Farmer, J. D., Smith, D. E., Shubik, M.. "Is Economics the Next Physical Science?" Physics Today 58(9) (2005): 37-42.
\item Foellmer, H., 1974. Random economies with many interacting agents. Journal of mathematical economics (1) pp 51 ff\par
\item Gulko, L., 1999. The entropic market hypothesis. International Journal of theoretical and applied finance, vol. 2, n0. 3, pp 293-329\par
\item Hildenbrand, W, Kirman (1988). Equilbrium Anaysis, North-Holland, Amsterdam\par
\item Kaldor, N. 1972. The Irrelevance of Equilibrium Economics. The Economic Journal, Vol. 82, no. 328, 1237-1255\par
\item Mantegna, R. N., Stanley, E., 2000. An Introduction to Econophysics: correlations and complexity in finance, Cambridge University Press\par
\item Paul, W., Baschnagel,J., 2000. Stochastic Processes. From Physics to Finance, Springer \par
\item Voit, J., 2005. The Statistical Mechanics of Financial Markets, Springer, Berlin \end{enumerate}}
 
\end{document}